\documentclass[a4paper,11pt]{article}
\usepackage{pos}
\usepackage{amsmath}
\usepackage{natbib}
\usepackage{aas_macros}
\newcommand{\vct}[1]{\mathrm{\mathbf{#1}}}
\newcommand{\uvct}[1]{\hat{\mathrm{\mathbf{#1}}}}

\usepackage{xcolor}

\title{Cosmic Ray Small Scale Anisotropies in Slab Turbulence}

\ShortTitle{Small Scale Anisotropies in Slab Turbulence}

\author*[a]{Marco Kuhlen}

\author[a]{Philipp Mertsch}

\author[a]{Vo Hong Minh Phan}

\affiliation[a]{RWTH-Aachen University, Institute for Theoretical Particle Physics and Cosmology (TTK)\\
Templergraben 55, 52056 Aachen, Germany}

\emailAdd{marco.kuhlen@rwth-aachen.de}
\emailAdd{pmertsch@physik.rwth-aachen.de}
\emailAdd{vhmphan@physik.rwth-aachen.de} 

\abstract{
In the standard picture of cosmic ray transport the propagation of charged
cosmic rays through turbulent magnetic fields is described as a random walk
with cosmic rays scattering on magnetic field turbulence. This is in good
agreement with the highly isotropic cosmic ray arrival directions as this
diffusion process effectively isotropizes the cosmic ray distribution.

High-statistics observatories like IceCube and HAWC have however observed significant
deviations from isotropy down to very small angular scales. This is in strong tension
with this standard picture of cosmic ray propagation. While large scale
multipoles arise naturally, for example due to the earth's motion relative to
the isotropic cosmic ray distribution, there is no intuitive mechanism to
account for the observed anisotropies at smaller angular scales. 

By relaxing one of the standard assumptions of quasi linear theory and treating
correlations between fluxes of cosmic rays from different directions explicitly
we show that higher multipoles also are to be expected from particle propagation
through turbulent magnetic fields. We present a first analytical calculation
of the angular power spectrum assuming a physically motivated model of the
magnetic field turbulence and find good agreement with numerical simulations.}

\FullConference{37$^{\rm{th}}$ International Cosmic Ray Conference (ICRC 2021)\\
		July 12th -- 23rd, 2021\\
		Online -- Berlin, Germany}

\begin{document}
\maketitle

\section{Introduction}
The arrival directions of Galactic cosmic rays (CRs) are expected to be highly isotropic due to the interaction of these particles with interstellar turbulent magnetic fields. 
This random scattering process effectively isotropizes their arrival directions and leads to the diffusive transport of CRs.
Current high-statistics observatories like IceCube~\cite{2019ApJ...871...96A} and HAWC~\cite{2018ApJ...865...57A} have however observed significant deviations from isotropy down to angular scales of $10^\circ$. These deviations from an isotropic CR arrival direction map can be quantified by the angular power spectrum (APS) defined as 
\begin{equation}
  C_\ell(t) = \frac{1}{4\pi}\int\mathrm{d}\uvct{p}_A\int\mathrm{d}\uvct{p}_B P_\ell(\uvct{p}_A\cdot\uvct{p}_B)f_A(\vec{r}_\oplus,\vct{p}_A,t)f_B(\vec{r}_\oplus,\vct{p}_B,t), \label{eqn:def_APS}
\end{equation}
where $\uvct{p}=\vct{p}/|\vct{p}|$ is a unit momentum vector and $f(\vct{r}_\oplus,\vct{p},t)$ is the phase-space density measured by an observer at position $\vct{r}_\oplus$ and time $t$ from direction $\uvct{p}$.

Interestingly, it has been suggested that the turbulent magnetic fields could be a potential source of small-scale anisotropies \cite{giacinti2012}. This idea has been investigated using test particle simulations of CRs in simulated turbulence. A map of the CR arrival directions can be obtained from the phase space density back-tracked along the cosmic ray trajectories to an earlier time (see \cite{2017PrPNP..94..184A} for a review). Most of the previous studies, however, limit themselves to particles with a ratio of Larmor radius to outer scale $\rho = r_g/L_\text{max} \sim 10^{-2}-10^{1}$. This corresponds to energies much larger than what is relevant for observational data, $\rho\lesssim10^{-3}$, assuming a typical value of the Galactic magnetic field~\cite{2015ApJ...815L...2A}. In that case, it might not be straightforward to make the direct comparison between the simulated and observed APS.

More importantly, the standard picture of CR transport known as quasi-linear theory (QLT) fails to explain the observed anisotropy at scales smaller than the dipole since it allows computing only the ensemble average of the phase-space density $\langle f\rangle$. This means that one has to assume $\langle f_Af_B\rangle = \langle f_A\rangle\langle f_B\rangle$ in order to calculate the ensemble averaged APS from Eq.~\eqref{eqn:def_APS}. 
While under these assumptions the ensemble averaged APS from standard QLT $\langle C_\ell\rangle^\mathrm{std}\sim0$ for $\ell\geq 2$ taking the correlations of phase-space densities into account correctly can lead to a larger APS since $\langle f_A f_B \rangle \geq \langle f_A\rangle \langle f_B \rangle$~\cite{2015ApJ...815L...2A}.
These correlations are expected to be present since particles arriving under an angle $\theta$ propagate through a single realization of the turbulent magnetic field. They will therefore experience similar magnetic fields. The authors of Ref. \cite{2019JCAP...11..048M} take into account these correlations and put forward a model to predict the APS based on a perturbative expansion of the time-evolution operator. However, a rather unrealistic white-noise power spectrum of turbulence was adopted to allow for some explicit analytical results.

The aim of this work is, thus, to further improve our understanding of turbulence induced small-scale anisotropies from both the simulational and analytical points of view. We will first present an extended version of the analytical framework to calculate the APS introduced in Ref.~\cite{2019JCAP...11..048M} for the case of slab turbulence.
The results are then compared to the simulated APS in the energy range relevant for observations for a given turbulence model.

\section{Analytical Calculation}
\label{sec:analytical}
We follow the steps outlined in Ref.~\cite{2019JCAP...11..048M}. For further details we refer to this paper and references therein.
The time evolution of the phase-space density $f$ is described by Liouville's equation 
\begin{equation}
  \partial_t f + \uvct{p}\nabla f + \mathcal{L} f = -\delta \mathcal{L} f \, ,
\end{equation}
where $\mathcal{L} = -i\vct{\Omega}\cdot L$ and $\delta\mathcal{L} = -i\omega\cdot\vct{L}$ are the relativistic and stochastic Liouville operators. Here $\vct{\Omega} = q \vct{B}_0/p_0$ and $\vct{\omega} = q\vct{\delta B}/p_0$ denote the gyrovectors in the regular and turbulent magnetic fields respectively and $\vct{L}$ is a vector of angular momentum operators that obey the usual commutation relations $[L_i,L_j]=i\epsilon_{ijk} L_k$.
After expanding the phase-space density around the position of the observer this equation can be solved formally by introducing a time evolution operator 
\begin{equation}
U_{t,t_0} = \mathcal{T}\exp{\left[-\int_{t_0}^{t}\mathrm{d}t'(\mathcal{L}+\delta\mathcal{L}(t'))\right]},
\end{equation}
with the time ordered exponential.

Similar to Feynman diagrams in quantum field theory the correlation of the time evolution operators can be expanded diagrammatically in the the strength of the turbulent magnetic field,
\begin{equation}
  \begin{matrix}\includegraphics[width=0.9\textwidth,trim={1cm 0cm 1cm 0cm}]{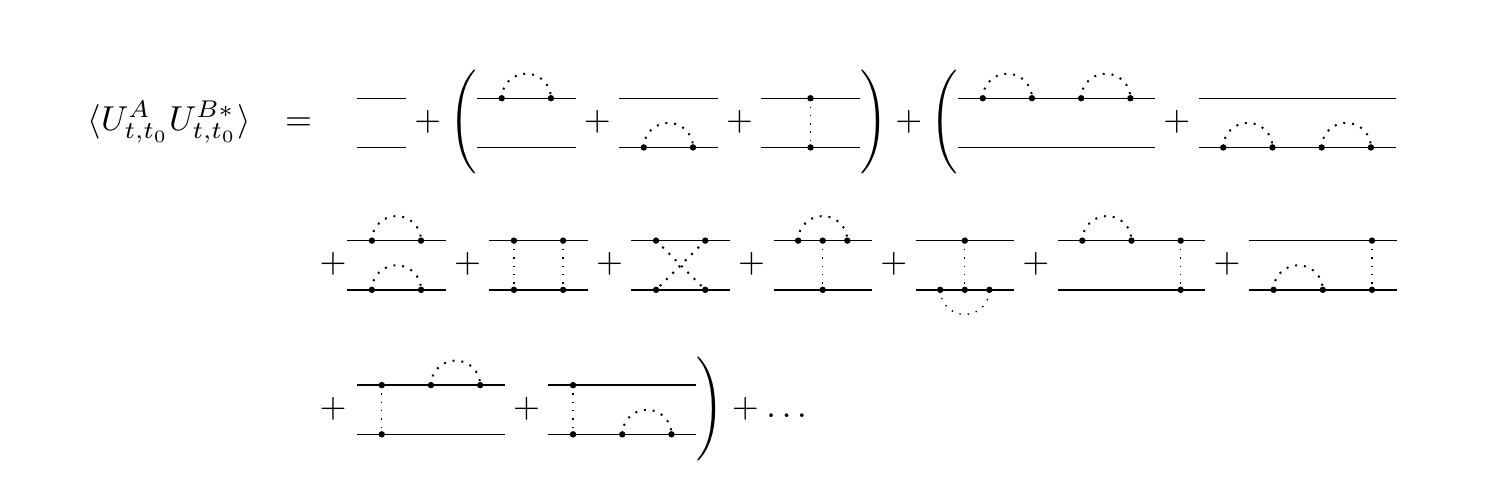}\end{matrix}
\end{equation}

We will limit ourselves to a first order calculation and only compute the diagrams up to and including the first parenthesis. With this correlation of the time evolution operators we can then define the mixing matrix
\begin{equation}
  M_{\ell\ell_0}(t,t_0) = \frac{1}{4\pi}\int\mathrm{d}\uvct{p}_A\int\mathrm{d}\uvct{p}_B P_\ell(\uvct{p}_A\cdot\uvct{p}_B)\langle U^A_{t,t_0}U^{B*}_{t,t_0}\rangle \frac{2\ell_0+1}{4\pi}P_{\ell_0}(\uvct{p}_A\cdot\uvct{p}_B),
\end{equation}
as the projection of $\langle U^A_{t,t_0}U^{B*}_{t,t_0}\rangle$ into the space of statistically isotropic $\langle f_A f_B\rangle$.
Making a gradient ansatz for the phase-space density a differential equation for the local time evolution of the angular power spectrum can be derived. The steady state angular power spectrum can be shown to satisfy
\begin{equation}
  \label{eq:mm}
  \frac{\delta_{\ell\ell_0}-M_{\ell\ell_0}(\Delta T)}{\Delta T} C_{\ell_0}^\mathrm{stdy}(t) = \frac{8\pi}{9}K|\nabla \bar{f}|^2\delta_{\ell 1},
\end{equation}
with the diffusion tensor $K$ and CR gradient $\nabla f$. The right hand side of this equation can be interpreted as a dipole term sourced by a CR gradient according to Fick's law. The left hand side describes how power from the dipole source is mixed into higher multipoles by the mixing matrix.

Assuming slab turbulence we find for the mixing matrix
\begin{equation}
\begin{aligned}
    \label{eq:mixingmatrix}
    M_{\ell\ell_0} &= \delta_{\ell\ell_0} - 8\pi\ell(\ell+1)\left(\frac{2}{3}\Lambda_0(\Delta T) - \frac{1}{3} \Lambda_2(\Delta T)\right)\delta_{\ell\ell_0} \\
    &+ 2\pi\sum_{\ell_A,\ell_B}i^{\ell_B-\ell_A}(2\ell_0+1){(2l_A+1)(2l_B+1)}\frac12     \begin{pmatrix}
      \ell_A & \ell & \ell_0 \\
      0 & 0 & 0 
    \end{pmatrix}
    \begin{pmatrix}
      \ell_B & \ell & \ell_0 \\
      0 & 0 & 0 
    \end{pmatrix}\\
    &\times(1+(-1)^{\ell_A+\ell_B})\sum_{m_0,m}\bigg(
    (2\ell_0(\ell_0+1)-2m_0^2)
    \begin{pmatrix}
      \ell_A & \ell & \ell_0 \\
      0 & m & m_{0}
    \end{pmatrix}
    \begin{pmatrix}
      \ell_B & \ell & \ell_0 \\
      0 & m & m_{0}
    \end{pmatrix}
    \kappa_{\ell_A,\ell_B}(\Delta T)\bigg)
\end{aligned}
\end{equation}
where $(:\,:\,:)$ denotes the Wigner 3j symbol and $\Lambda_i(\Delta T)$ and $\kappa_i(\Delta T)$ are integrals defined as
\begin{equation}
  \Lambda_{\ell_A}(\Delta T) = \int_0^{\Delta T}\mathrm{d}T\int_0^T\mathrm{d}\tau\int\mathrm{d}k~ g(k) \cos{(\Omega\tau)} j_{\ell_A}(k\tau)
\end{equation}
and
\begin{equation}
  \kappa_{\ell_A,\ell_B}(\Delta T) = \int_{t_0}^t\mathrm{d}t_1\int_{t_0}^t\mathrm{d}t_2\int\mathrm{d}k~
  g(k)j_{\ell_A}(k(t-t_1))j_{\ell_B}(k(t-t_2))\cos(\Omega (t_1-t_2)),
\end{equation}
involving the power spectrum of magnetic turbulence $g(k)$, defined as the magnitude dependent part of the Fourier transform of the magnetic two point correlation function $g(k)(\delta_{ij}+k_ik_j/k^2) = 1/(2\pi)^3\int\mathrm{d}^3x\exp{(-i\vec{k}\cdot\vec{x})}\langle\delta B_i(\vec{x}_0)\delta B_j(\vec{x}_0+\vec{x})\rangle$ and the spherical Bessel functions $j_\ell(x)$.

The terms in the first line of eq.~\eqref{eq:mixingmatrix} describe the pitchangle scattering known from QLT. Since this part of the mixing matrix is diagonal they do not lead to mixing between different multipoles. If it was for these to contributions only, the steady-state APS would be only dipolar. The third term in eq.~\eqref{eq:mixingmatrix} comes from including the correlations and this is what leads to a non-vanishing angular power spectrum at larger multipoles $\ell$.

\begin{figure}
  \centering
  \includegraphics[width=0.7\textwidth]{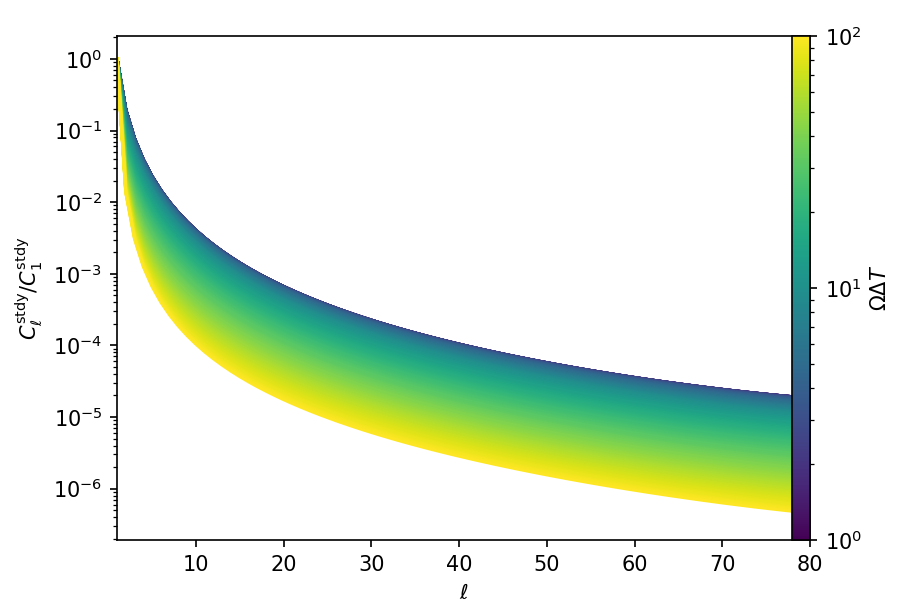}
  \caption{Angular power spectra from the analytical calculation normalised to the dipole as a function of $\ell$. The value of the parameter $\Omega \Delta T$ is colour-coded.}
  \label{fig:analytic_results}
\end{figure}

Putting these contributions to the mixing matrix together Eq.~\eqref{eq:mm} can be solved numerically to get the steady state angular power spectrum. The resulting angular power spectrum for different values of the remaining model parameter $\Omega\Delta T$ is shown in fig.~\ref{fig:analytic_results}.

\section{Numerical Simulation}
To verify the results of our analytical calculation we use numerical simulations of testparticles in synthetic turbulence~\cite{2020Ap&SS.365..135M}. The monoenergetic testparticles are initialized at the origin with isotropic directions on a HEALPix~\cite{2005ApJ...622..759G}  grid with $N_\text{side} = 256~\mathrm{or}~512$ leading to a total number of testparticles $N_\text{particles}\sim 8\times 10^5~\text{and}~3\times 10^6$. These particles are then tracked back in time through the magnetic field by solving the Newton-Lorentz equations using the energy conserving Boris method~\cite{boris}. 
As the testparticles do not interact with each other or backreact on the magnetic field this can be parallelized very efficiently. We therefore run these simulations on GPUs which allow for efficient parallelisation.
For all simulations we choose a maximal wavelength $L_\text{max} = 150\,\text{pc}$~\cite{2008ApJ...680..362H} and a total root mean square magnetic field strength $\sqrt{B_0^2+\langle\delta B^2\rangle} = 4\,\mu G$~\cite{2003astro.ph.10287B}. The turbulence level $\eta = \langle\delta B^2\rangle/(B_0^2+\langle\delta B^2\rangle)$ is varied between $0.1$ and $0.5$~\cite{2003astro.ph.10287B}.

In the literature two different methods have been used to generate synthetic magnetic field turbulence. In the method proposed by Giacalone and Jokipii~\cite{1994ApJ...430L.137G}, the magnetic field is calculated as a superposition of waves. Only the phases and amplitudes for the waves are stored. In the other method the magnetic field is set up on a grid in Fourier space, transformed to and saved in real space. This has the advantage that no large sums have to be evaluated at every particle position. The magnetic field evaluation is reduced to a simple interpolation between grid points. The disadvantage is the large amount of memory required to store the entire field grid.

\begin{figure}
  \centering
  \includegraphics[width=\textwidth]{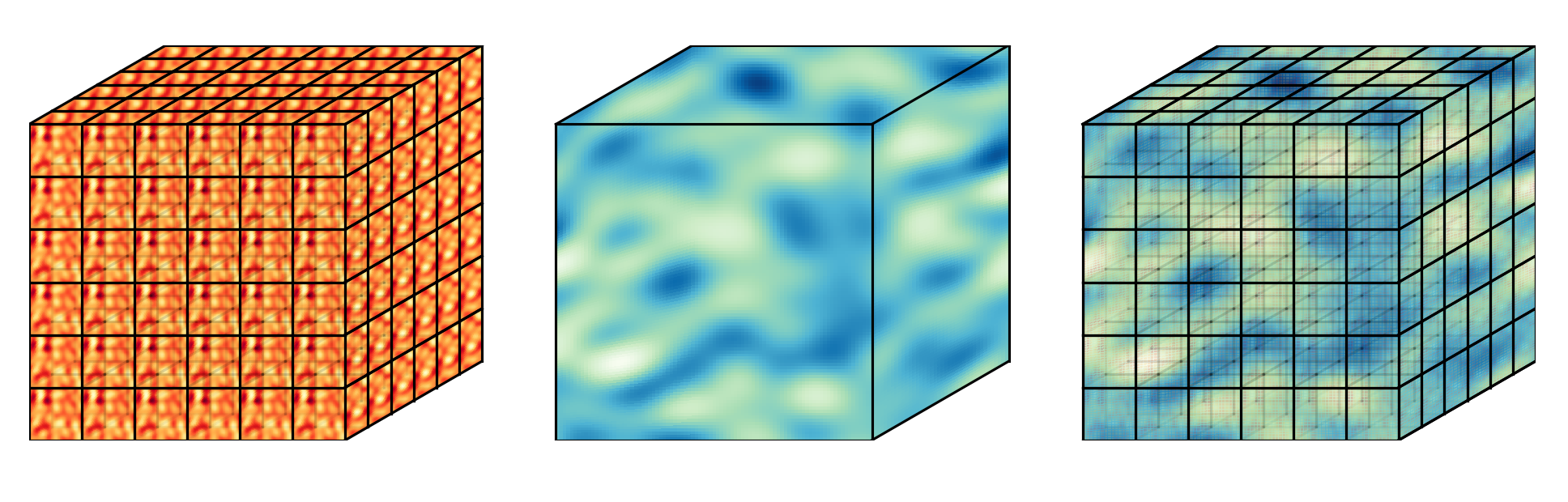}
  \caption{Illustration of the nested grid method. The cube on the left shows the small grid used to resolve the short wavelengths. It is then replicated multiple times to span the simulation volume. The cube in the centre shows the large grid used to resolve the long wavelenghts. The right cube is given by the superposition of the two cubes and contains both long and short wavelengths.}
  \label{fig:nested_cubes}
\end{figure}

The rigidity of particles required to compare to observational data from IceCube and HAWC is of the order of $10\,\text{TV}$. The resulting gyroradii and thus also the minimal wavelength that needs to be resolved in our simulations is therefore $r_g\sim 2.5\times 10^{-3}\,\text{pc}$. Even though there could be artefacts due to grid periodicity, it has proven sufficient to make the grid a factor of  $8$ larger than the largest wavelength. The smallest wavelength is chosen a factor $20$ smaller than the gyroradius of the particles and resolved by at least $10$ grid points. Spanning this large dynamical range with a single grid would require at least $n = 96,000,000$  grid points. To reduce the memory requirement on our GPUs we therefore use 3 nested grids    with different grid spacings as proposed by~\cite{2012JCAP...07..031G}. Each individual grid covers a part of the magnetic field power spectrum. The 3 grids are then superimposed. An illustration of the nested grid method for 2 grids without padding is shown in Fig.~\ref{fig:nested_cubes}. 

In section~\ref{sec:analytical} the angular power spectrum was calculated by evolving an assumed initial phase-space density forward in time using a differential equation derived from the Vlasov equation. To calculate the angular power spectrum from the backtracked particle trajectories the local phase-space density at time $t$ $f(\mathbf{r}=0, \mathbf{p},t)$ is related to the phase-space density at an earlier time $t-T$ along a CR trajectory $i$, with coordinate $\mathbf{r}_i(t)$ and momentum $\mathbf{p}_i(t)$ using Liouville's theorem
\begin{equation}
  f_i(\mathbf{r}_i=0,\mathbf{p}_i,t) = f(\mathbf{r}_i(t-T),\mathbf{p}_i(t-T),t-T).
\end{equation}
Assuming a quasi stationary solution to the diffusion equation $\langle f_i(t-T)\rangle  \simeq \bar{f}-3\hat{\mathbf{p}}_i\cdot\mathbf{K}\cdot\nabla\bar{f}$ and only small          fluctuations from the ensemble average $\delta f = f-\langle f \rangle$ the phase-space density at $t$ is given by
\begin{equation}
  f_i(t) \simeq \delta f(\mathbf{r}_i(t-T),\mathbf{p}_i(t-T),t-T) + \bar{f} + \mathbf{r_i}(t-   T)\cdot\nabla\bar{f} - 3\hat{\mathbf{p}}_i(t-T)\cdot \mathbf{K}\cdot \nabla \bar{f}.
\end{equation}
The two methods of calculating the angular power spectrum are of course equivalent.
The angular power spectrum calculated from the numerical simulations converges to a constant  angular power spectrum for large times. This is due to the angular power spectrum being sensitive only to the realization of the \emph{local} magnetic  field.

Even for a large number of particles the higher multipoles of the angular power spectrum are strongly effected by shot noise. For large backtracking times it can be estimated as~\cite{2015ApJ...815L...2A}
\begin{equation}
  \mathcal{N} = \frac{4\pi}{N_\text{pix}}2TK_{ij}\frac{\partial_i                          \bar{f}\partial_j\bar{f}}{\bar{f}^2},
\end{equation}
where $N_\text{pix}$ is the number of pixels in the arrival direction skymap which in this case is equal to the number of particles in the simulation.
The noise-subtracted power spectrum can be estimated~\cite{2015MNRAS.448.2854C} via $\hat{C_\ell}=\langle C_\ell\rangle -\mathcal{N}$ with the variance  $\sigma^2(\hat{C_\ell})\simeq 2\mathcal{N}^2/(2\ell+1)$.

\section{Results}
The steady state APS shown in Fig.~\ref{fig:analytic_results} exhibit a power law-like falling behaviour in $\ell$ with a slope that depends on the parameter $\Omega \Delta T$. For large $\Omega \Delta T$ the power spectrum drops off faster than for small $\Omega \Delta T$. 

This parameter can be interpreted as the number of gyrotimes over which correlations in the 1c diagram decay. In QLT the particles trajectories are approximated as unperturbed trajectories. The particles can thus interact with the waves for an infinitely long time leading to a sharp resonance. In reality however particles trajectories are perturbed by the interactions with the turbulent field leading to a decay of correlations on a timescale related to the scattering time $\tau_s$.

\begin{figure}
  \centering
  \includegraphics[width=0.7\textwidth]{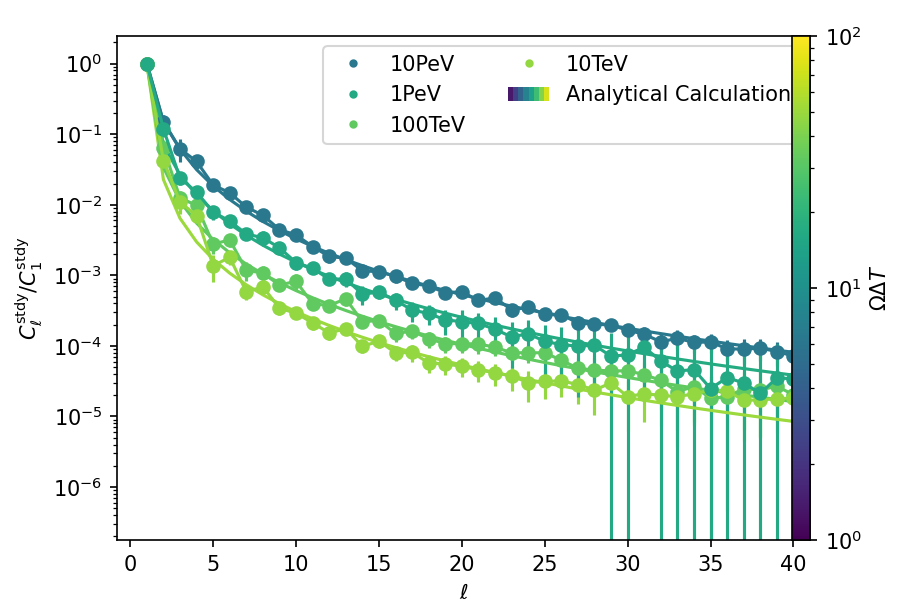}
  \caption{Angular power spectra of the arrival directions of CRs. The points show the angular power spectra computed from the results of test particle simulations in slab turbulence with a turbulence level of $\eta = 0.5$. Also shown is the best fit line from the analytical results for each individual energy. The numerical and analytical results agree well down to the smallest angular scales.}
  \label{fig:fit}
\end{figure}

The angular power spectra from the numerical simulations are shown in fig.~\ref{fig:fit}. They also fall like a powerlaw in $\ell$ with the slope being larger for smaller energies. Also shown are the best fit lines from the set of analytical solutions.

From the resonance structure of the contribution of the interacting diagram we expect the $\Omega\Delta T$ parameter to be related to the scattering time via $\Omega\Delta T\propto (\Omega\tau_s)^{1/3}$. This relation is confirmed by the best fit $\Omega\Delta T$. Measuring the angular power spectrum thus gives an independent handle on $\Omega\tau_s$.

\section{Conclusion}
The angular power spectrum of CR arrival directions is an important observable. If it is indeed sourced by correlations of particles experiencing the same turbulent magnetic field it can be used to understand the local field configuration. 
Understanding the origin of the angular power spectrum is therefore important as an independent probe of the outer scale of turbulence and the local turbulence geometry.

Here we have used a perturbative calculation to predict the angular power spectrum of CR arrival directions taking into account the correlations of phase space densities implied by the correlations in the turbulent magnetic field. We have assumed a homogeneous background magnetic field such that the unperturbed orbits are helical. The perturbative expansion up to first order in the turbulence strength then includeds resonance effects between particles and the turbulent magnetic field similarly to QLT. 
The difference in the angular power spectrum compared to QLT arises because also correlations between phase-space densities that are induced by particles travelling through the same field realization are treated explicitly. This leads to a finite angular power spectrum even at larger multipoles $\ell$.

To validate and test the assumptions that were made in this calculation we have compared to testparticle simulations done in the same turbulence model at the rigidities relevant for observations by IceCube and HAWC. 
This comparison shows very good agreement between the analytical model and the numerical testparticle simulations. 

\bibliographystyle{JHEP}
\bibliography{Bibliography}

\end{document}